\documentclass[a4paper,11pt]{article}
\usepackage{pos}
\usepackage{orcidlink}
\usepackage{graphicx}
\usepackage{dcolumn}
\usepackage{bm}
\usepackage{textgreek}
\usepackage{placeins}
\usepackage{braket}
\usepackage{physics}
\usepackage{overpic}
\usepackage{amsmath}
\usepackage{subfigure}
\usepackage{qcircuit}
\usepackage{tikz}

\title{Quantum Simulation of Large $N_c$ Lattice Gauge Theories}

\author*[a]{Anthony N. Ciavarella \,\orcidlink{0000-0003-3918-4110}}
\author[a,b]{Christian W. Bauer \,\orcidlink{0000-0001-9820-5810}}

\affiliation[a]{Physics Division, Lawrence Berkeley National Laboratory, \\
Berkeley, California 94720, USA}

\affiliation[b]{Department of Physics, University of California, \\
Berkeley, Berkeley, CA 94720}

\emailAdd{anciavarella@lbl.gov}
\emailAdd{cwbauer@lbl.gov}

\abstract{A Hamiltonian lattice formulation of lattice gauge theories opens the possibility for quantum simulations of the non-perturbative dynamics of QCD. By parametrizing the gauge invariant Hilbert space in terms of plaquette degrees of freedom, we show how the Hilbert space and interactions can be expanded in inverse powers of $N_c$. At leading order in this expansion, the Hamiltonian simplifies dramatically, both in the required size of the Hilbert space as well as the type of interactions involved. Adding a truncation of the resulting Hilbert space in terms of local electric energy states we give explicit constructions that allow simple representations of SU(3) gauge fields on qubits and qutrits to leading order in large $N_c$.}

\FullConference{%
  The 41st International Symposium on Lattice Field Theory, \\
  28 July 2024 \\
  University of Liverpool, Liverpool, United Kingdom
}

\begin{document}
\maketitle

\section{Introduction}
Quantum computers offer the potential to explore the dynamics of strongly coupled gauge theories with computational resources that scale polynomially in the system size~\cite{feynman2018simulating,Bauer_2023,humble2022snowmass,humble2022snowmass2,beck2023quantum,dimeglio2023quantum,nielsen2001quantum}. This is anticipated to enable first principle simulations of the inelastic scattering of hadrons, calculations of the QCD shear viscosity and other observables that involve real-time dynamics~\cite{humble2022snowmass,humble2022snowmass2,beck2023quantum,dimeglio2023quantum,moore2020shear}. The development of the first quantum computers has enabled exploratory studies of one-dimensional systems and small two-dimensional systems~\cite{martinez2016real,klco20202,Rahman:2022rlg,rahman2022real,ciavarella2021trailhead,ciavarella2022preparation,alam2022primitive,Illa:2022jqb,Gustafson_2022,Atas_2021,farrell2023preparations,farrell2023preparations2,Atas:2022dqm,yang2020observation,Zhou_2022,Su_2023,zhang2023observation,mildenberger2022probing,ciavarella2023quantum,farrell2023scalable,Farrell:2024fit,charles2023simulating,kavaki2024square,mueller2022quantum}. The quantum simulation of systems with multiple spacial dimensions has been limited by the complexity of implementing plaquette operators. This has motivated searches for more practical encodings of gauge fields onto quantum hardware.

In the study of $SU(N_c)$ gauge theories, it has proven fruitful to work in the large $N_c$ limit and expand in powers of $1 / N_c$~\cite{tHooft:1973alw,Sjostrand:2006za,Bahr:2008pv,PICH_2002,KAPLAN1996244}. In this limit, the theory simplifies dramatically. This limit provides a starting point for a description of meson interactions and is an essential ingredient in the parton shower approximation~\cite{LUCINI201393,Manohar:1998xv,Sjostrand:2006za,Bahr:2008pv}. In this work, it will be shown how to combine a large $N_c$ expansion with the Hamiltonian formulation of lattice gauge theories. This enables simplified encodings of the gauge fields onto discrete degrees of freedom which can be used in near-term quantum simulations.

\section{Large $N_c$ Counting}
In the Hamiltonian formulation of $SU(N_c)$ lattice Yang-Mills, each link on the lattice has a Hilbert space spanned by states of the form $\ket{U}$ where $U\in SU(N_c)$. Alternatively, the Hilbert space can be described in terms of irreducible representations with basis states $\ket{j,m_L,m_R}$ where $j$ specifies an irreducible representation, and $m_{L/R}$ represents a component of that representation associated with the left (right) side of that link. This irrep, also known as electric, basis is discrete which makes it suitable for use in quantum simulation. In this basis, the electric energy operator on each link is diagonal,
\begin{equation}
     \hat{E}^2 \ket{j,m_L,m_R} = C(j) \ket{j,m_L,m_R} \ \ \ ,
\end{equation}
where $C(j)$ is the Casimir of the representation $j$ and the matrix elements of the parallel transporters are given by
\begin{equation}
   \bra{j',m'_L,m'_R} \hat{U}_{\alpha \beta} \ket{j,m_L,m_R} = \sqrt{\frac{\text{dim}(j)}{\text{dim}(j')}} C^{j' m'_L}_{j m_L; F \alpha} C^{j' m'_R}_{j m_R; F \beta} \ \ \ ,
\end{equation}
where $C^{A a}_{B b; C c}$ is the Clebsch-Gordan coefficient for the representation $A$ being formed by combining representations $B$ and $C$, and $F$ is the fundamental representation of the gauge group. Gauge invariant operators can be constructed by taking products of parallel transporters around some closed curve, i.e.,
\begin{equation}
    \hat{U}_C = \prod_{l\in C} \sum_{s_i} \hat{U}^{l}_{s_i, s_{i+1}} \ \ \ ,
\end{equation}
where $l$ are the links along the curve $C$. The Kogut-Susskind Hamiltonian~\cite{kogut1975hamiltonian,kogut1979introduction,banks1977strong,jones1979lattice} for Hamiltonian lattice gauge theory is given by
\begin{equation}
    \hat{H} = \sum_l \frac{g^2}{2} \hat{E}^2_l - \frac{1}{2g^2} \sum_{p \in \text{plaquettes}} \left(\hat{\Box}_p + \hat{\Box}^\dagger_p\right) \ \ \ ,
    \label{eq:LQCDFull}
\end{equation}
where $g$ is the strong coupling constant, $\hat{E}_{l}^2$ is the electric energy operator on link $l$ and $\Box_p$ is the product of parallel transporters on plaquette $p$. Note that the following discussion will focus on $2+1D$, but the techniques should work in any spacetime dimension. 

To determine what electric basis states need to be represented on a quantum computer to simulate dynamics, we can consider generating the vacuum state by adiabatically turning on the plaquette terms in the Hamiltonian. Excited states in the simplest topological sector can be obtained by further applications of electric energy or plaquette operators to the state. From this approach, it can be seen that the only physically relevant electric basis states are those with non-zero overlap on states of the form 
\begin{align}
	\ket{\{P_p, \bar P_p\}}\equiv \prod_p \hat{\Box}_p^{P_p} \hat{\Box}_p^{\dagger \bar P_p} \ket{0} 
	\ \ \ ,
\end{align}
where $\ket{0}$ is the electric vacuum state. One can therefore classify all states in this topological sector by the minimum number of plaquette operators and its conjugate that are required to reach it from the vacuum. Therefore, the large $N_c$ scaling of different basis states can be determined by the maximal overlap of the basis state with states of the form $\ket{\{P_p, \bar P_p\}}\equiv \prod_p \hat{\Box}_p^{P_p} \hat{\Box}_p^{\dagger \bar P_p} \ket{0}$.

\begin{figure}
    \centering    \includegraphics[width=0.85\textwidth]{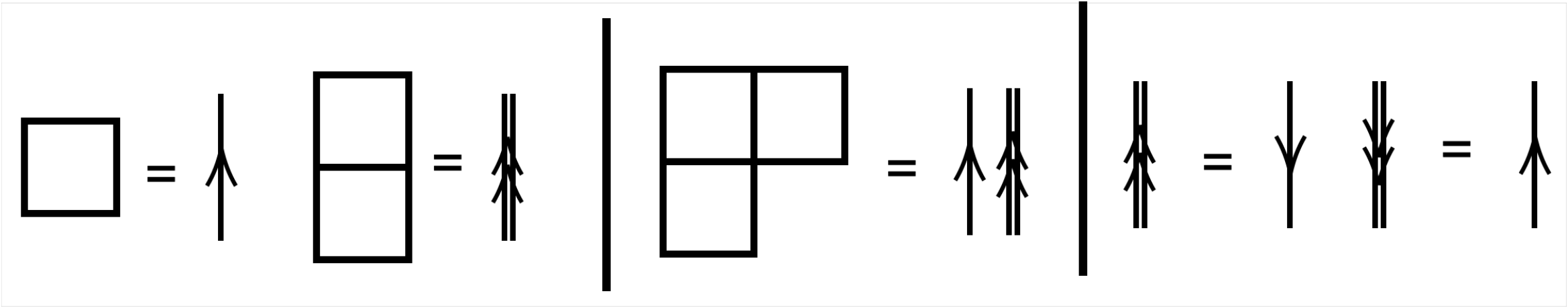}
    \caption{Graphical representations of Young diagrams in terms of arrows.}
    \label{fig:arrowRep}
\end{figure}

\begin{figure}
    \centering 
    \includegraphics[width=6.6cm]{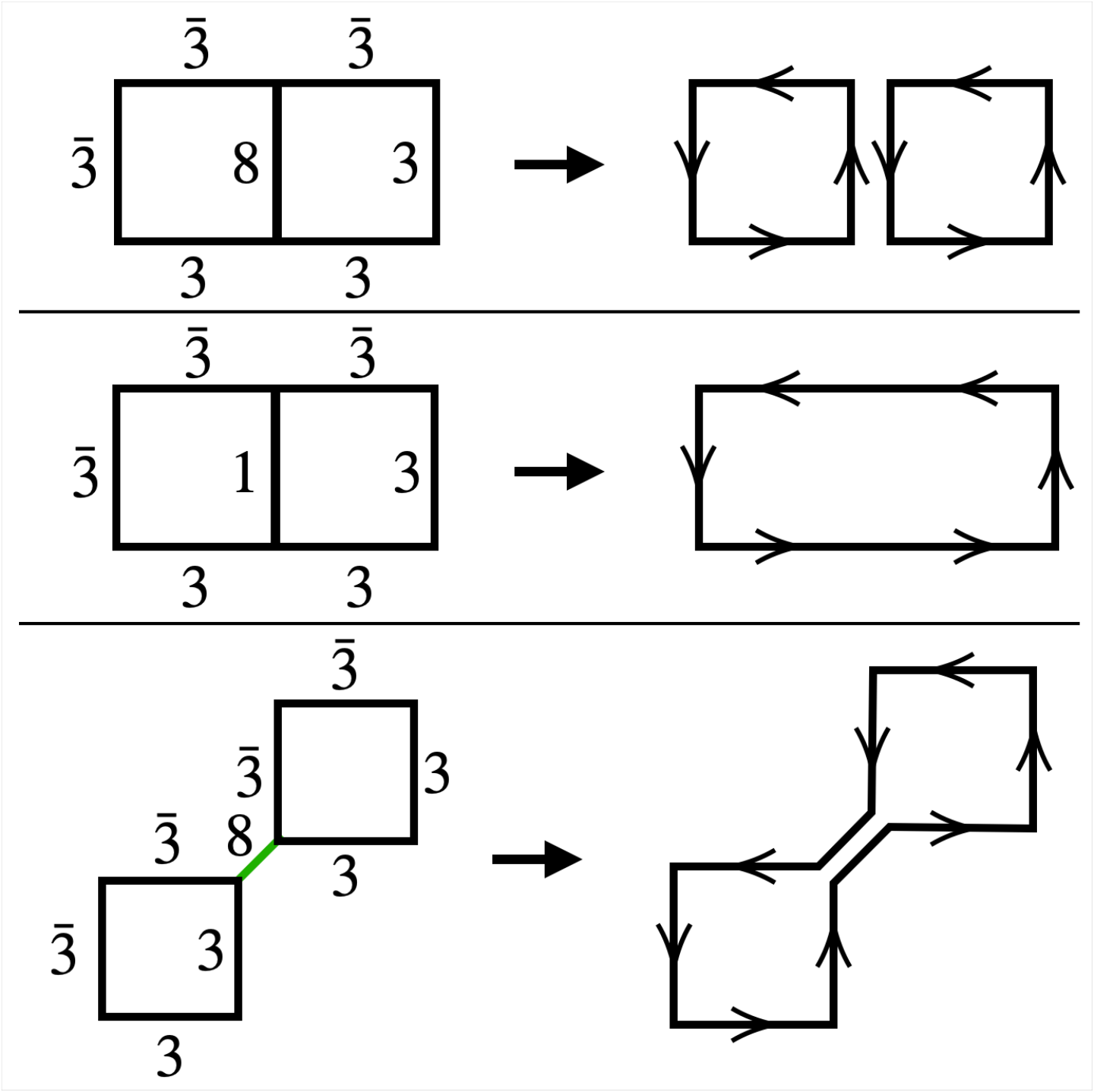}
    \caption{Graphical representations of basis states on a point-split lattice.}
    \label{fig:repRelation}
\end{figure}

To aid in the large $N_c$ counting, a graphical notation for gauge invariant states will be introduced. Physical states are subject to a constraint from Gauss's law which requires that the sum of representations on each vertex of the lattice forms a singlet. On a lattice where each vertex is connected to at most three links, gauge invariant states can be specified by the representation $R$ on each link and a specification on each vertex of how the links add to form a singlet. For $SU(N_c)$ gauge groups, a representation $R$ can be labeled by a Young diagram, which can be specified through the number of columns with $1, 2, \ldots, N-1$ boxes. For $SU(3)$ only two numbers are required, and the labels are often chosen as $(p, q)$, with $p$ labeling the number of columns with a single box, and $q$ labeling the number of columns with two boxes. Young diagrams can be represented by lines with arrows, as illustrated in Fig.~\ref{fig:arrowRep}. 
One can see that fundamental and anti-fundamental representations can be represented either by lines with a single arrow in one direction or by lines with a double arrow in the opposite direction. More complex representations can be built by combining such lines together. For lattices where vertices connect to more links, such as a square lattice in 2D or 3D, not all states that can be labeled by the representation above are linearly independent, leading to an ambiguity in labeling the basis states. This is due to the so-called Mandelstam constraints, which relate contractions of representation indices across a vertex. A point-splitting procedure can be performed to split each vertex into three link vertices connected by virtual links, which lifts this ambiguity. In this point-split lattice, the gauge-invariant states can be specified with the same assignment of labels used on a trivalent lattice. There is an equivalent labeling of the states of the physical Hilbert space that will prove useful, using the arrow representation introduced above. This is illustrated in Fig.~\ref{fig:repRelation}, and will be called a ``loop representation''
In this representation, each state is labeled by a set of loops $L_i$, together with a specification $a_\ell$, which denotes the way the arrows at each link $\ell$ having more than one loop pass through being combined. Each loop needs to specify which plaquettes are encircled and in which order, while $a_\ell$ contains the information on how to combine lines of multiple loops into single or double arrows. Note that this loop representation is simply a graphical representation of the states with definite representation at each link.

With this graphical notation, large $N_c$ scaling rules can be derived for basis states $\ket{\{L_i, a_\ell\}}$. The large $N_c$ scaling of a state, $\ket{\{L_i, a_\ell\}}$ is determined by the large $N_c$ expansion of $\braket{\{L_i, a_\ell\}}{\{P_p, \bar P_p\}}$ for the minimal choice of $P_p$ and $\bar P_p$ to obtain a nonzero overlap. Defining $\ket{\{L_i\}} = \prod_i U_{L_i} \ket{0}$ where $U_{L_i}$ is a product of parallel transporters along the loop $L_i$ and using that the overlap $\braket{\{L_i, a_\ell\}}{\{L_i\}}$ is ${\cal O}(1)$ in the $N_c$ scaling, the $N_c$ scaling is determined by the overlap $\braket{\{L_i\}}{\{P_p, \bar P_p\}}$.This overlap can be evaluated in the magnetic basis through inserting $\mathbf{1} = \prod_{\text{links l}} \int dU_l \ket{U_l} \bra{U}_l$. To evaluate the large $N_c$ scaling of these integrals, the identity 
\begin{align}
    & \int dU \prod_{n=1}^q U_{i_n j_n} U^{*}_{i'_n j'_n} = \nonumber \\
    & \frac{1}{N_c^q} \sum_{\text{permutations k}} \prod_{n=1}^q \delta_{i_n i'_{k_n}} \delta_{j_n j'_{k_n}} + \mathcal{O}\left(\frac{1}{N_c^{q+1}}\right) \,,
    \label{eq:Uint}
\end{align}
will be used~\cite{weingarten1978asymptotic}.
\begin{figure*}
    \centering    \includegraphics[width=.9\textwidth]{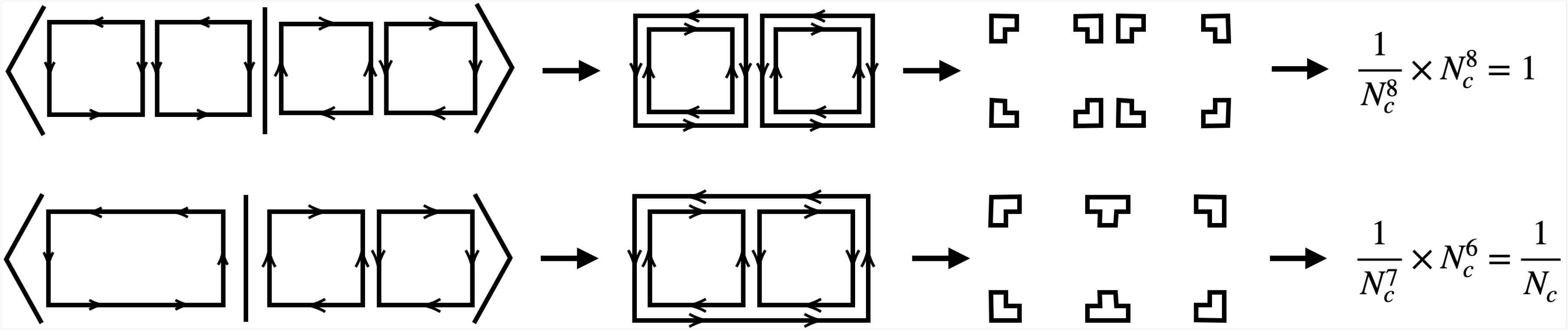}
    \caption{Graphical method to obtain the scaling of the overlap matrix $\braket{\{L_i\}}{\{P_p, \bar P_p\}}$. The top example contains two loops, each encircling a single plaquette $m_1 = m_2 = 1$, while the bottom example has a single loop encircling 2 loops $m_1 = 1$. This gives for the top example $q_1 = q_2 = 1+3\times 1 = 4$ and $v_1 = v_2 = 2 + 2\times 1 = 4$, giving the final scaling $N_c^0$. For the bottom example we have $q_1 = 1 + 2 \times 3 = 7$ and $v_1 = 2 + 2\times 2 = 6$, giving the final scaling $1/N_c$. }
    \label{fig:NScale}
\end{figure*}
The large $N_c$ scaling will be determined by the permutation of indices contraction that gives the largest factors of $N_c$. 
A diagrammatic method of evaluating the large $N_c$ scaling is shown in Fig.~\ref{fig:NScale}. 

First, the plaquette operators being applied are placed over loops in the final state. 
To determine the powers of $N_c$ that come from contracting the Kronecker $\delta$s, one can erase the middle of each link in the diagram and connect the lines from the same vertex.
This leaves a set of $v$ closed loops involving one vertex each, and each of these closed loops contributes a factor of $N_c$ in the numerator. 
Each loop $L_i$ therefore contributes a factor $N_c^{v_i - q_i}$ and the the total $N_c$ scaling is given by
\begin{align}
    N_c^{v - q}\,, \qquad  q\equiv \sum_i q_i\,, \quad v\equiv \sum_i v_i
\end{align}
to the final overlap.
Since each $U_{ij}$ in Eq.~\eqref{eq:Uint} corresponds to a line in the figure, one immediately finds that $q = n_l/2$, where $n_l$ is the total number of lines on each link in the diagram. 
Denoting by $m_i$ the number of plaquettes encircled by each loop $L_i$, one needs $m_i$ plaquette operators for each loop. 
The total number of lines is then given by $n_l = 2 + 6m_i$, and the total number of closed loops $n_v$ is given by $2 + 2 m_i$ for each loop in the basis. Thus one finds
\begin{align}
    q_i = 1 + 3 m_i \,, \qquad v_i = 2 + 2m_i
    \,.
\end{align}
Putting this together, one finds that each loop contributes a factor of $N_c^{1 - m_i}$
to the overall scaling of the overlap, such that
\begin{align}
    \braket{\{L_i, a_\ell\}}{\{P_p, \bar P_p\}} \propto \prod_i N_c^{1 - m_i}
    \,,
\end{align}
This implies that the states that can be reached to leading order in $1/N_c$ are those that only involve loops $L_i$ with $m_i = 1$.
Therefore, the only overlap that survives in the large $N_c$ limit is the one with states $\ket{\{L_i, a_\ell\}}$ for which each loop encircles exactly one plaquette. At order $1/N_c$, states with loops extending over two plaquettes will be present.

\section{Truncated Hamiltonians}
\subsection{Large $N_c$ Limit}
With the large $N_c$ scaling determined, it is possible to develop truncations of the Kogut-Susskind Hamiltonian that both limit the electric energy per link and truncate states at some order in the large $N_c$ expansion. The harshest possible truncation keeps only the fundamental and anti-fundamental representations ($\mathbf{3}$ and $\mathbf{\Bar{3}}$ for $SU(3)$) and only allows loops around single plaquettes. The Hilbert space at this truncation can therefore be described by assigning a qutrit to each plaquette in the lattice. The states of the qutrit will be labelled by $\ket{0}$, $\ket{\circlearrowleft}$, and $\ket{\circlearrowright}$. Physical states are subject to the constraint that neighboring plaquettes are not simultaneously excited. For example, in a two plaquette system, the states $\ket{\circlearrowleft}\ket{0}$ and $\ket{\circlearrowright}\ket{0}$ are physical while $\ket{\circlearrowright}\ket{\circlearrowright}$ and $\ket{\circlearrowleft} \ket{\circlearrowright}$ are not, since it would give rise to the common link having $p+q > 1$.
 The electric field operator for a link $\ell$ lying on plaquettes $p$ and $p'$ at this truncation can be written as
\begin{align}
    \hat{E_\ell}^2 = & \frac{4}{3}\left[ \ket{\circlearrowleft}_{p}\bra{\circlearrowleft}_{p}+ \ket{\circlearrowleft}_{p'}\bra{\circlearrowleft}_{p'}+ \left(\ket{\circlearrowleft} \leftrightarrow  \ket{\circlearrowright}\right)\right] \,,
\end{align}
where we have used the full expression of the Casimir of the fundamental representation $C_f = (N_c^2 - 1) / (2 N_c) = 4/3$.
The plaquette operator at position $p$ is given by
\begin{align}
    \hat{\Box}_{p} = &  \hat{P}_{0,p+\hat x} \hat{P}_{0,p-\hat x} \hat{P}_{0,p+\hat y} \hat{P}_{0,p-\hat y} \nonumber \\
    & \times \left(\ket{\circlearrowleft}_p \bra{0}_p + \ket{\circlearrowright}_p \bra{\circlearrowleft}_p + \ket{0}_p \bra{\circlearrowright}_p\right) \,,
\end{align}
where $\hat{P}_{0,p} = \ket{0}_p \bra{0}_p$ and $p\pm\hat x$ ($p\pm\hat y$) denotes the plaquette one position away in the $x$ ($y$) direction. 

This Hamiltonian has a charge conjugation (C) symmetry that causes states with the anti-symmetric combination $\frac{1}{\sqrt{2}}\left(\ket{\circlearrowleft} - \ket{\circlearrowright}\right)$ anywhere on the lattice to decouple from the rest of the Hilbert space. 
This decoupling can be seen by repeated applications of the plaquette operators to the electric vacuum. Explicitly, we have
\begin{align}
    \hat{\Box}_{p} & \ket{0} = \ket{\circlearrowleft} + \ket{\circlearrowright} \nonumber \\
    \hat{\Box}_{p} & \frac{1}{\sqrt{2}}\left(\ket{\circlearrowleft} + \ket{\circlearrowright}\right) = \sqrt{2} \ket{0} + \frac{1}{\sqrt{2}}\left(\ket{\circlearrowleft} + \ket{\circlearrowright} \right)  \,,
\end{align}
so the state $\frac{1}{\sqrt{2}}\left(\ket{\circlearrowleft} - \ket{\circlearrowright}\right)$ is never coupled to the rest of the Hilbert space.
One can therefore perform separate simulations for the C even and odd sector.
By assigning $\ket{1} = \frac{1}{\sqrt{2}}\left(\ket{\circlearrowleft} \pm \ket{\circlearrowright}\right)$, the C (anti)symmetric subspace can be described by assigning a qubit to each plaquette instead of a qutrit. 
As already mentioned, physical states have the constraint that neighboring qubits cannot both be in the $\ket{1}$ state. 
With this encoding, the Hamiltonian for the C even sector is given by
\begin{align}
    \hat{H} =& \sum_p \left(\frac{8}{3}g^2 - \frac{1}{2g^2}\right) \hat{P}_{1,p} \nonumber \\
    & - \frac{1}{g^2\sqrt{2}}\hat{P}_{0,p+\hat x} \hat{P}_{0,p-\hat x} \hat{P}_{0,p+\hat y} \hat{P}_{0,p-\hat y} \hat{X}_p \,,
    \label{eq:Ham3PXP}
\end{align}
where $\hat{P}_{1,p} = \ket{1}_p \bra{1}_p$ and $\hat{X}_p$ is the Pauli X operator acting on the qubit at plaquette $p$. Note that the coefficient of $\frac{1}{2g^2}$ multiplying $\hat{P}_{1,p}$ is a somewhat unique feature of $SU(3)$ as only in $SU(3)$ will two applications of the plaquette operator to the vacuum state produce the anti-fundamental representation.

\subsection{Subleading in $1/N_c$}
\begin{figure}
    \centering
    \includegraphics[width=0.5\linewidth]{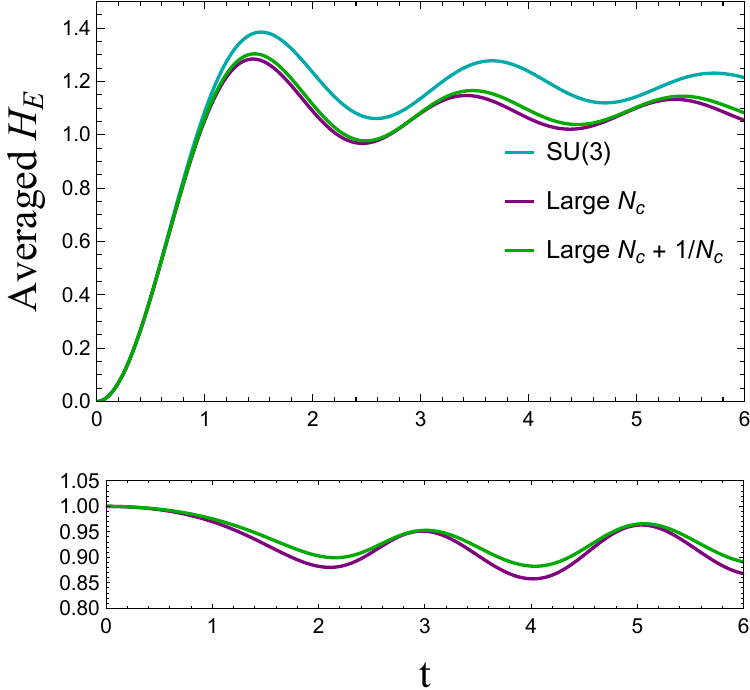}
    \caption{Calculation of $\frac{1}{T}\int_0^T dt \bra{\psi(t)} \hat{H}_E \ket{ \psi(t)}$ on a $4\times1$ lattice with periodic boundary conditions and $g=1$. The blue line shows the simulation for a SU(3) lattice gauge theory truncated at $p+q\leq1$, using the formalism introduced in Ref~\cite{ciavarella2021trailhead}. The purple line shows the time evolution computed with the large $N_c$ truncated Hamiltonian in Eq.~\eqref{eq:Ham3PXP}. The green line shows the time evolution with the $1/N_c$ corrections from Eq.~\eqref{eq:Ham1N}. The lines underneath show the ratio of the large $N_c$ electric energy to the SU(3) electric energy.}
    \label{fig:SU3LargeNComparison}
\end{figure}

To obtain a more accurate description of finite $N_c$ gauge theories, one needs to go beyond leading order in the large $N_c$ expansion. As discussed before, at order $1 / N_c$ terms in the plaquette operator that extend loops to neighboring plaquettes must be included. If the electric truncation is kept at $p+q \leq 1$, the theory at this truncation can once again be represented using a qutrit per plaquette. However, now neighboring qutrits in the states $\ket{\circlearrowleft} \ket{\circlearrowleft}$ or $\ket{\circlearrowright} \ket{\circlearrowright}$ will be interpreted as having their shared link in the $\mathbf{1}$ irrep. At this truncation, it is also possible to project into the $C$ even sector, and use a single qubit per plaquette. In this basis, a region of qubits in the $\ket{1}$ state is interpreted as having a loop of electric flux in an even superposition of $\mathbf{3}$ and $\mathbf{\Bar{3}}$ flowing around the boundary of the region. With this basis, the Hamiltonian is given by
\begin{align}
    \hat{H}_{1/N} =& \sum_p \frac{8}{3}g^2  \hat{P}_{1,p} - \frac{4}{3}g^2 \hat{P}_{1,p}\left( \hat{P}_{1,p +\hat{x}} + \hat{P}_{1,p -\hat{x}} + \hat{P}_{1,p +\hat{y}} + \hat{P}_{1,p -\hat{y}} \right) \nonumber \\
    & - \frac{1}{g^2\sqrt{2}}\hat{P}_{0,p+\hat x} \hat{P}_{0,p-\hat x} \hat{P}_{0,p+\hat y} \hat{P}_{0,p-\hat y} \hat{X}_p - \frac{1}{2g^2}\hat{P}_{0,p+\hat x} \hat{P}_{0,p-\hat x} \hat{P}_{0,p+\hat y} \hat{P}_{0,p-\hat y}  \hat{P}_{1,p} \nonumber \\
    & - \frac{1}{6g^2} \hat{X}_p \left( \hat{P}_{1,p+\hat x} \hat{P}_{0,p-\hat x} \hat{P}_{0,p+\hat y} \hat{P}_{0,p-\hat y} + \hat{P}_{0,p+\hat x} \hat{P}_{1,p-\hat x} \hat{P}_{0,p+\hat y} \hat{P}_{0,p-\hat y}\right) \nonumber \\
    & - \frac{1}{6g^2} \hat{X}_p \left( \hat{P}_{0,p+\hat x} \hat{P}_{0,p-\hat x} \hat{P}_{1,p+\hat y} \hat{P}_{0,p-\hat y} + \hat{P}_{0,p+\hat x} \hat{P}_{0,p-\hat x} \hat{P}_{0,p+\hat y} \hat{P}_{1,p-\hat y}\right) \ \ \ .
    \label{eq:Ham1N}
\end{align}
To probe the effects of large $N_c$ truncations, the electric vacuum was evolved in time on a $4\times1$ lattice with PBC. Fig.~\ref{fig:SU3LargeNComparison} shows the evolution of $\frac{1}{T} \int_0^T dt \bra{\psi(t)} \hat{H}_E \ket{ \psi(t)}$ as a function of $T$ for a SU(3) LGT truncated at $p+q\leq1$. At long times, this observable is expected to equilibrate to a thermal value determined by the initial state's energy. As Fig.~\ref{fig:SU3LargeNComparison} shows, the relative error from the large $N_c$ expansion at leading order is roughly $20\%$ which should be expected from expanding in $\frac{1}{N_c}$ with $N_c=3$. Including the $1 / N_c$ corrections gives modest improvements. Note that the improvements are not as large as one would expect because the full simulation includes vertices with three incoming $\mathbf{3}$ irreps which in the large $N_c$ approach only show up when higher irrep states are included.

\section{Summary}
In this work, large $N_c$ expansions have been combined with the Hamiltonian formulation of lattice gauge theories to reduce the resources required to map the theory onto a quantum computer. The lowest-lying truncations require only a single qubit per plaquette. Remarkably, the lowest-lying truncation takes the form of a $PXP$ model which is known to display quantum scarring~\cite{nandkishore2015many,choi2019emergent,Moudgalya_2022,chandran2023quantum,surace2020lattice}. This suggests similar phenomena may also occur in $SU(3)$ gauge theory. This model is also a limit of the Ising model with transverse and longitudinal fields which displays confinement~\cite{kormos2017real,james2019nonthermal,robinson2019signatures}. This suggests that connecting confinement in the Ising model to the physics of large $N_c$ Yang-Mills may be possible. It is expected that the formalism introduced here can be extended to higher spacial dimensions and include matter. The simplifications obtained by truncating in powers of $1/N_c$ may enable near-term quantum simulations of physically relevant phenomena such as inelastic scattering or jet fragmentation in lattice QCD.

\acknowledgments
This contribution is based on the work in Ref~\cite{ciavarella2024quantum} written by the same authors. We would like to acknowledge helpful conversations with Ivan Burbano, Irian D'Andrea, Jesse Stryker, and Michael Kreshchuk. We would like to thank Martin Savage, Marc Illa, and Roland Farrell for many conversations related to quantum simulation. We would like to thank Aneesh Manohar for helpful discussions about large $N_c$ expansions. We would also like to acknowledge helpful conversations with Jad Halimeh about the emergence of $PXP$ models from certain limits of gauge theories. 
This material is based 
upon work supported by the U.S. Department of Energy, Office of Science, National Quantum Information 
Science Research Centers, Quantum Systems Accelerator. Additional support is acknowledged from the U.S. Department of Energy (DOE), Office of Science under contract DE-AC02-05CH11231, partially through Quantum Information Science Enabled Discovery (QuantISED) for High Energy Physics (KA2401032).

\bibliographystyle{JHEP}
\bibliography{ref}

\providecommand{\href}[2]{#2}\begingroup\raggedright\begin{thebibliography}{10}

\bibitem{feynman2018simulating}
R.P.~Feynman, \emph{{Simulating physics with computers}}, \href{https://doi.org/10.1007/BF02650179}{\emph{Int. J. Theor. Phys.} {\bfseries 21} (1982) 467}.

\bibitem{Bauer_2023}
C.W.~Bauer et~al., \emph{{Quantum Simulation for High-Energy Physics}}, \href{https://doi.org/10.1103/PRXQuantum.4.027001}{\emph{PRX Quantum} {\bfseries 4} (2023) 027001} [\href{https://arxiv.org/abs/2204.03381}{{\ttfamily 2204.03381}}].

\bibitem{humble2022snowmass}
T.S.~Humble et~al., \emph{{Snowmass White Paper: Quantum Computing Systems and Software for High-energy Physics Research}},  in \emph{{Snowmass 2021}}, 3, 2022 [\href{https://arxiv.org/abs/2203.07091}{{\ttfamily 2203.07091}}].

\bibitem{humble2022snowmass2}
T.S.~Humble, G.N.~Perdue and M.J.~Savage, \emph{{Snowmass Computational Frontier: Topical Group Report on Quantum Computing}},  \href{https://arxiv.org/abs/2209.06786}{{\ttfamily 2209.06786}}.

\bibitem{beck2023quantum}
D.~Beck et~al., \emph{{Quantum Information Science and Technology for Nuclear Physics. Input into U.S. Long-Range Planning, 2023}},  2, 2023 [\href{https://arxiv.org/abs/2303.00113}{{\ttfamily 2303.00113}}].

\bibitem{dimeglio2023quantum}
A.~Di~Meglio et~al., \emph{{Quantum Computing for High-Energy Physics: State of the Art and Challenges}}, \href{https://doi.org/10.1103/PRXQuantum.5.037001}{\emph{PRX Quantum} {\bfseries 5} (2024) 037001} [\href{https://arxiv.org/abs/2307.03236}{{\ttfamily 2307.03236}}].

\bibitem{nielsen2001quantum}
M.A.~Nielsen and I.L.~Chuang, \emph{Quantum Computation and Quantum Information: 10th Anniversary Edition}, Cambridge University Press, New York, NY, USA, 10th~ed. (2011).

\bibitem{moore2020shear}
G.D.~Moore, \emph{{Shear viscosity in QCD and why it's hard to calculate}},  in \emph{{Criticality in QCD and the Hadron Resonance Gas}}, 10, 2020 [\href{https://arxiv.org/abs/2010.15704}{{\ttfamily 2010.15704}}].

\bibitem{martinez2016real}
E.A.~Martinez et~al., \emph{{Real-time dynamics of lattice gauge theories with a few-qubit quantum computer}}, \href{https://doi.org/10.1038/nature18318}{\emph{Nature} {\bfseries 534} (2016) 516} [\href{https://arxiv.org/abs/1605.04570}{{\ttfamily 1605.04570}}].

\bibitem{klco20202}
N.~Klco, J.R.~Stryker and M.J.~Savage, \emph{{SU(2) non-Abelian gauge field theory in one dimension on digital quantum computers}}, \href{https://doi.org/10.1103/PhysRevD.101.074512}{\emph{Phys. Rev. D} {\bfseries 101} (2020) 074512} [\href{https://arxiv.org/abs/1908.06935}{{\ttfamily 1908.06935}}].

\bibitem{Rahman:2022rlg}
S.~A~Rahman, R.~Lewis, E.~Mendicelli and S.~Powell, \emph{{Self-mitigating Trotter circuits for SU(2) lattice gauge theory on a quantum computer}}, \href{https://doi.org/10.1103/PhysRevD.106.074502}{\emph{Phys. Rev. D} {\bfseries 106} (2022) 074502} [\href{https://arxiv.org/abs/2205.09247}{{\ttfamily 2205.09247}}].

\bibitem{rahman2022real}
E.~Mendicelli, R.~Lewis, S.A.~Rahman and S.~Powell, \emph{{Real time evolution and a traveling excitation in SU(2) pure gauge theory on a quantum computer.}}, \href{https://doi.org/10.22323/1.430.0025}{\emph{PoS} {\bfseries LATTICE2022} (2023) 025} [\href{https://arxiv.org/abs/2210.11606}{{\ttfamily 2210.11606}}].

\bibitem{ciavarella2021trailhead}
A.~Ciavarella, N.~Klco and M.J.~Savage, \emph{{Trailhead for quantum simulation of SU(3) Yang-Mills lattice gauge theory in the local multiplet basis}}, \href{https://doi.org/10.1103/PhysRevD.103.094501}{\emph{Phys. Rev. D} {\bfseries 103} (2021) 094501} [\href{https://arxiv.org/abs/2101.10227}{{\ttfamily 2101.10227}}].

\bibitem{ciavarella2022preparation}
A.N.~Ciavarella and I.A.~Chernyshev, \emph{{Preparation of the SU(3) lattice Yang-Mills vacuum with variational quantum methods}}, \href{https://doi.org/10.1103/PhysRevD.105.074504}{\emph{Phys. Rev. D} {\bfseries 105} (2022) 074504} [\href{https://arxiv.org/abs/2112.09083}{{\ttfamily 2112.09083}}].

\bibitem{alam2022primitive}
{\scshape SQMS} collaboration, \emph{{Primitive quantum gates for dihedral gauge theories}}, \href{https://doi.org/10.1103/PhysRevD.105.114501}{\emph{Phys. Rev. D} {\bfseries 105} (2022) 114501} [\href{https://arxiv.org/abs/2108.13305}{{\ttfamily 2108.13305}}].

\bibitem{Illa:2022jqb}
M.~Illa and M.J.~Savage, \emph{{Basic Elements for Simulations of Standard Model Physics with Quantum Annealers: Multigrid and Clock States}}, \href{https://doi.org/10.1103/PhysRevA.106.052605}{\emph{Phys. Rev. A} {\bfseries 106} (2022) 052605} [\href{https://arxiv.org/abs/2202.12340}{{\ttfamily 2202.12340}}].

\bibitem{Gustafson_2022}
E.J.~Gustafson, H.~Lamm, F.~Lovelace and D.~Musk, \emph{{Primitive quantum gates for an SU(2) discrete subgroup: Binary tetrahedral}}, \href{https://doi.org/10.1103/PhysRevD.106.114501}{\emph{Phys. Rev. D} {\bfseries 106} (2022) 114501} [\href{https://arxiv.org/abs/2208.12309}{{\ttfamily 2208.12309}}].

\bibitem{Atas_2021}
Y.Y.~Atas, J.~Zhang, R.~Lewis, A.~Jahanpour, J.F.~Haase and C.A.~Muschik, \emph{{SU(2) hadrons on a quantum computer via a variational approach}}, \href{https://doi.org/10.1038/s41467-021-26825-4}{\emph{Nature Commun.} {\bfseries 12} (2021) 6499} [\href{https://arxiv.org/abs/2102.08920}{{\ttfamily 2102.08920}}].

\bibitem{farrell2023preparations}
R.C.~Farrell, I.A.~Chernyshev, S.J.M.~Powell, N.A.~Zemlevskiy, M.~Illa and M.J.~Savage, \emph{{Preparations for quantum simulations of quantum chromodynamics in 1+1 dimensions. I. Axial gauge}}, \href{https://doi.org/10.1103/PhysRevD.107.054512}{\emph{Phys. Rev. D} {\bfseries 107} (2023) 054512} [\href{https://arxiv.org/abs/2207.01731}{{\ttfamily 2207.01731}}].

\bibitem{farrell2023preparations2}
R.C.~Farrell, I.A.~Chernyshev, S.J.M.~Powell, N.A.~Zemlevskiy, M.~Illa and M.J.~Savage, \emph{{Preparations for quantum simulations of quantum chromodynamics in 1+1 dimensions. II. Single-baryon \ensuremath{\beta}-decay in real time}}, \href{https://doi.org/10.1103/PhysRevD.107.054513}{\emph{Phys. Rev. D} {\bfseries 107} (2023) 054513} [\href{https://arxiv.org/abs/2209.10781}{{\ttfamily 2209.10781}}].

\bibitem{Atas:2022dqm}
Y.Y.~Atas, J.F.~Haase, J.~Zhang, V.~Wei, S.M.L.~Pfaendler, R.~Lewis et~al., \emph{{Simulating one-dimensional quantum chromodynamics on a quantum computer: Real-time evolutions of tetra- and pentaquarks}}, \href{https://doi.org/10.1103/PhysRevResearch.5.033184}{\emph{Phys. Rev. Res.} {\bfseries 5} (2023) 033184} [\href{https://arxiv.org/abs/2207.03473}{{\ttfamily 2207.03473}}].

\bibitem{yang2020observation}
B.~Yang, H.~Sun, R.~Ott, H.-Y.~Wang, T.V.~Zache, J.C.~Halimeh et~al., \emph{{Observation of gauge invariance in a 71-site Bose\textendash{}Hubbard quantum simulator}}, \href{https://doi.org/10.1038/s41586-020-2910-8}{\emph{Nature} {\bfseries 587} (2020) 392} [\href{https://arxiv.org/abs/2003.08945}{{\ttfamily 2003.08945}}].

\bibitem{Zhou_2022}
Z.-Y.~Zhou, G.-X.~Su, J.C.~Halimeh, R.~Ott, H.~Sun, P.~Hauke et~al., \emph{{Thermalization dynamics of a gauge theory on a quantum simulator}}, \href{https://doi.org/10.1126/science.abl6277}{\emph{Science} {\bfseries 377} (2022) abl6277} [\href{https://arxiv.org/abs/2107.13563}{{\ttfamily 2107.13563}}].

\bibitem{Su_2023}
G.-X.~Su, H.~Sun, A.~Hudomal, J.-Y.~Desaules, Z.-Y.~Zhou, B.~Yang et~al., \emph{{Observation of many-body scarring in a Bose-Hubbard quantum simulator}}, \href{https://doi.org/10.1103/PhysRevResearch.5.023010}{\emph{Phys. Rev. Res.} {\bfseries 5} (2023) 023010} [\href{https://arxiv.org/abs/2201.00821}{{\ttfamily 2201.00821}}].

\bibitem{zhang2023observation}
W.-Y.~Zhang et~al., \emph{{Observation of microscopic confinement dynamics by a tunable topological $\theta$-angle}},  \href{https://arxiv.org/abs/2306.11794}{{\ttfamily 2306.11794}}.

\bibitem{mildenberger2022probing}
J.~Mildenberger, W.~Mruczkiewicz, J.C.~Halimeh, Z.~Jiang and P.~Hauke, \emph{{Probing confinement in a $\mathbb{Z}_2$ lattice gauge theory on a quantum computer}},  \href{https://arxiv.org/abs/2203.08905}{{\ttfamily 2203.08905}}.

\bibitem{ciavarella2023quantum}
A.N.~Ciavarella, \emph{{Quantum simulation of lattice QCD with improved Hamiltonians}}, \href{https://doi.org/10.1103/PhysRevD.108.094513}{\emph{Phys. Rev. D} {\bfseries 108} (2023) 094513} [\href{https://arxiv.org/abs/2307.05593}{{\ttfamily 2307.05593}}].

\bibitem{farrell2023scalable}
R.C.~Farrell, M.~Illa, A.N.~Ciavarella and M.J.~Savage, \emph{{Scalable Circuits for Preparing Ground States on Digital Quantum Computers: The Schwinger Model Vacuum on 100 Qubits}}, \href{https://doi.org/10.1103/PRXQuantum.5.020315}{\emph{PRX Quantum} {\bfseries 5} (2024) 020315} [\href{https://arxiv.org/abs/2308.04481}{{\ttfamily 2308.04481}}].

\bibitem{Farrell:2024fit}
R.C.~Farrell, M.~Illa, A.N.~Ciavarella and M.J.~Savage, \emph{{Quantum simulations of hadron dynamics in the Schwinger model using 112 qubits}}, \href{https://doi.org/10.1103/PhysRevD.109.114510}{\emph{Phys. Rev. D} {\bfseries 109} (2024) 114510} [\href{https://arxiv.org/abs/2401.08044}{{\ttfamily 2401.08044}}].

\bibitem{charles2023simulating}
C.~Charles, E.J.~Gustafson, E.~Hardt, F.~Herren, N.~Hogan, H.~Lamm et~al., \emph{{Simulating Z2 lattice gauge theory on a quantum computer}}, \href{https://doi.org/10.1103/PhysRevE.109.015307}{\emph{Phys. Rev. E} {\bfseries 109} (2024) 015307} [\href{https://arxiv.org/abs/2305.02361}{{\ttfamily 2305.02361}}].

\bibitem{kavaki2024square}
A.H.Z.~Kavaki and R.~Lewis, \emph{{From square plaquettes to triamond lattices for SU(2) gauge theory}}, \href{https://doi.org/10.1038/s42005-024-01697-4}{\emph{Commun. Phys.} {\bfseries 7} (2024) 208} [\href{https://arxiv.org/abs/2401.14570}{{\ttfamily 2401.14570}}].

\bibitem{mueller2022quantum}
N.~Mueller, J.A.~Carolan, A.~Connelly, Z.~Davoudi, E.F.~Dumitrescu and K.~Yeter-Aydeniz, \emph{{Quantum Computation of Dynamical Quantum Phase Transitions and Entanglement Tomography in a Lattice Gauge Theory}}, \href{https://doi.org/10.1103/PRXQuantum.4.030323}{\emph{PRX Quantum} {\bfseries 4} (2023) 030323} [\href{https://arxiv.org/abs/2210.03089}{{\ttfamily 2210.03089}}].

\bibitem{tHooft:1973alw}
G.~'t~Hooft, \emph{{A Planar Diagram Theory for Strong Interactions}}, \href{https://doi.org/10.1016/0550-3213(74)90154-0}{\emph{Nucl. Phys. B} {\bfseries 72} (1974) 461}.

\bibitem{Sjostrand:2006za}
T.~Sjostrand, S.~Mrenna and P.Z.~Skands, \emph{{PYTHIA 6.4 Physics and Manual}}, \href{https://doi.org/10.1088/1126-6708/2006/05/026}{\emph{JHEP} {\bfseries 05} (2006) 026} [\href{https://arxiv.org/abs/hep-ph/0603175}{{\ttfamily hep-ph/0603175}}].

\bibitem{Bahr:2008pv}
M.~Bahr et~al., \emph{{Herwig++ Physics and Manual}}, \href{https://doi.org/10.1140/epjc/s10052-008-0798-9}{\emph{Eur. Phys. J. C} {\bfseries 58} (2008) 639} [\href{https://arxiv.org/abs/0803.0883}{{\ttfamily 0803.0883}}].

\bibitem{PICH_2002}
A.~Pich, \emph{{Colorless mesons in a polychromatic world}}, \href{https://doi.org/10.1142/9789812776914_0023}{\emph{{The Phenomenology of Large N(c) QCD}} (2002) 239} [\href{https://arxiv.org/abs/hep-ph/0205030}{{\ttfamily hep-ph/0205030}}].

\bibitem{KAPLAN1996244}
D.B.~Kaplan and M.J.~Savage, \emph{{The Spin flavor dependence of nuclear forces from large n QCD}}, \href{https://doi.org/10.1016/0370-2693(95)01277-X}{\emph{Phys. Lett. B} {\bfseries 365} (1996) 244} [\href{https://arxiv.org/abs/hep-ph/9509371}{{\ttfamily hep-ph/9509371}}].

\bibitem{LUCINI201393}
B.~Lucini and M.~Panero, \emph{{SU(N) gauge theories at large N}}, \href{https://doi.org/10.1016/j.physrep.2013.01.001}{\emph{Phys. Rept.} {\bfseries 526} (2013) 93} [\href{https://arxiv.org/abs/1210.4997}{{\ttfamily 1210.4997}}].

\bibitem{Manohar:1998xv}
A.V.~Manohar, \emph{{Large N QCD}},  in \emph{{Les Houches Summer School in Theoretical Physics, Session 68: Probing the Standard Model of Particle Interactions}}, pp.~1091--1169, 2, 1998 [\href{https://arxiv.org/abs/hep-ph/9802419}{{\ttfamily hep-ph/9802419}}].

\bibitem{kogut1975hamiltonian}
J.~Kogut and L.~Susskind, \emph{Hamiltonian formulation of {W}ilson's lattice gauge theories}, \href{https://doi.org/10.1103/PhysRevD.11.395}{\emph{Phys. Rev. D} {\bfseries 11} (1975) 395}.

\bibitem{kogut1979introduction}
J.B.~Kogut, \emph{An introduction to lattice gauge theory and spin systems}, \href{https://doi.org/10.1103/RevModPhys.51.659}{\emph{Rev. Mod. Phys.} {\bfseries 51} (1979) 659}.

\bibitem{banks1977strong}
T.~Banks, S.~Raby, L.~Susskind, J.~Kogut, D.R.T.~Jones, P.N.~Scharbach et~al., \emph{Strong-coupling calculations of the hadron spectrum of {Q}uantum {C}hromodynamics}, \href{https://doi.org/10.1103/PhysRevD.15.1111}{\emph{Phys. Rev. D} {\bfseries 15} (1977) 1111}.

\bibitem{jones1979lattice}
D.R.T.~Jones, R.D.~Kenway, J.B.~Kogut and D.K.~Sinclair, \emph{{Lattice Gauge Theory Calculations Using an Improved Strong Coupling Expansion and Matrix Pade Approximants}}, \href{https://doi.org/10.1016/0550-3213(79)90190-1}{\emph{Nucl. Phys. B} {\bfseries 158} (1979) 102}.

\bibitem{weingarten1978asymptotic}
D.~Weingarten, \emph{{Asymptotic Behavior of Group Integrals in the Limit of Infinite Rank}}, \href{https://doi.org/10.1063/1.523807}{\emph{J. Math. Phys.} {\bfseries 19} (1978) 999}.

\bibitem{nandkishore2015many}
R.~Nandkishore and D.A.~Huse, \emph{{Many body localization and thermalization in quantum statistical mechanics}}, \href{https://doi.org/10.1146/annurev-conmatphys-031214-014726}{\emph{Ann. Rev. Condensed Matter Phys.} {\bfseries 6} (2015) 15} [\href{https://arxiv.org/abs/1404.0686}{{\ttfamily 1404.0686}}].

\bibitem{choi2019emergent}
S.~Choi, C.J.~Turner, H.~Pichler, W.W.~Ho, A.A.~Michailidis, Z.~Papi\'c et~al., \emph{{Emergent SU(2) Dynamics and Perfect Quantum Many-Body Scars}}, \href{https://doi.org/10.1103/PhysRevLett.122.220603}{\emph{Phys. Rev. Lett.} {\bfseries 122} (2019) 220603}.

\bibitem{Moudgalya_2022}
S.~Moudgalya, B.A.~Bernevig and N.~Regnault, \emph{{Quantum many-body scars and Hilbert space fragmentation: a review of exact results}}, \href{https://doi.org/10.1088/1361-6633/ac73a0}{\emph{Rept. Prog. Phys.} {\bfseries 85} (2022) 086501} [\href{https://arxiv.org/abs/2109.00548}{{\ttfamily 2109.00548}}].

\bibitem{chandran2023quantum}
A.~Chandran, T.~Iadecola, V.~Khemani and R.~Moessner, \emph{{Quantum Many-Body Scars: A Quasiparticle Perspective}}, \href{https://doi.org/10.1146/annurev-conmatphys-031620-101617}{\emph{Ann. Rev. Condensed Matter Phys.} {\bfseries 14} (2023) 443} [\href{https://arxiv.org/abs/2206.11528}{{\ttfamily 2206.11528}}].

\bibitem{surace2020lattice}
F.M.~Surace, P.P.~Mazza, G.~Giudici, A.~Lerose, A.~Gambassi and M.~Dalmonte, \emph{{Lattice gauge theories and string dynamics in Rydberg atom quantum simulators}}, \href{https://doi.org/10.1103/PhysRevX.10.021041}{\emph{Phys. Rev. X} {\bfseries 10} (2020) 021041} [\href{https://arxiv.org/abs/1902.09551}{{\ttfamily 1902.09551}}].

\bibitem{kormos2017real}
M.~Kormos, M.~Collura, G.~Tak\'acs and P.~Calabrese, \emph{{Real-time confinement following a quantum quench to a non-integrable model}}, \href{https://doi.org/10.1038/nphys3934}{\emph{Nature Phys.} {\bfseries 13} (2016) 246}.

\bibitem{james2019nonthermal}
A.J.A.~James, R.M.~Konik and N.J.~Robinson, \emph{{Nonthermal states arising from confinement in one and two dimensions}}, \href{https://doi.org/10.1103/PhysRevLett.122.130603}{\emph{Phys. Rev. Lett.} {\bfseries 122} (2019) 130603} [\href{https://arxiv.org/abs/1804.09990}{{\ttfamily 1804.09990}}].

\bibitem{robinson2019signatures}
N.J.~Robinson, A.J.A.~James and R.M.~Konik, \emph{{Signatures of rare states and thermalization in a theory with confinement}}, \href{https://doi.org/10.1103/PhysRevB.99.195108}{\emph{Phys. Rev. B} {\bfseries 99} (2019) 195108} [\href{https://arxiv.org/abs/1808.10782}{{\ttfamily 1808.10782}}].

\bibitem{ciavarella2024quantum}
A.N.~Ciavarella and C.W.~Bauer, \emph{{Quantum Simulation of SU(3) Lattice Yang-Mills Theory at Leading Order in Large-Nc Expansion}}, \href{https://doi.org/10.1103/PhysRevLett.133.111901}{\emph{Phys. Rev. Lett.} {\bfseries 133} (2024) 111901} [\href{https://arxiv.org/abs/2402.10265}{{\ttfamily 2402.10265}}].

\end{thebibliography}\endgroup

\end{document}